\documentclass[twocolumn,prl,superscriptaddress,floatfix,showpacs]{revtex4-1}
\usepackage[utf8]{inputenc}
\usepackage{amsmath}
\usepackage{amssymb}
\usepackage{graphicx}

\makeatletter
\usepackage{graphics}
\usepackage{epsfig}
\usepackage{color}

\newcommand{\be}{\begin{equation}} \newcommand{\ee}{\end{equation}}
\newcommand{\bea}{\begin{eqnarray}} \newcommand{\eea}{\end{eqnarray}}

\makeatother

\begin{document}

\title{Asymmetry of cross correlations between intra-day and overnight volatilities}

\author{Rubina Zadourian}

\affiliation{Max-Planck-Institut f\"ur Physik Komplexer Systeme, N\"othnitzer Strasse
38, 01187 Dresden, Germany}

\author{Peter Grassberger}

\affiliation{JSC, FZ J\"ulich, D-52425 J\"ulich, Germany}

\date{\today}
\begin{abstract}
We point out a stunning time asymmetry in the short time cross correlations
between intra-day and overnight volatilities (absolute values of log-returns
of stock prices). While overnight volatility is significantly (and
positively) correlated with the intra-day volatility during the \textit{following}
day (allowing thus non-trivial predictions), it is much less correlated 
with the intra-day volatility during the \textit{preceding}
day. While the effect is not unexpected in view of previous observations,
its robustness and extreme simplicity are remarkable. 
\end{abstract}
\maketitle
It is well known that fluctuations in equity prices traded on any
stock exchange are barely correlated, and hardly allow any non-trivial
forecasting. This is different for the \textit{amplitudes} of these
fluctuations, called \textit{volatilities}. If the market is hectic,
volatilities are large, and it will take some time until the market
has become calm again. With some modern options it is possible to
make profits with forecasts for volatilities (although forecasts of
signed fluctuations would be more easy to turn into money, if they
were possible), whence volatilities have been studied extensively
in the econometric literature \cite{bouchaud}.

As first shown in \cite{french,lockwood}, the statistics of volatilities
is not uniform. Rather, there is a marked daily structure, with high
volatility during the opening hour of the market and a more calm period
around noon. This seems to be true for all stock exchanges worldwide.
Also, equity prices at the opening of the trading session are in general
different from the closing prices on the previous trading day, showing
that there is a non-trivial overnight dynamics.

Finally, it has been repeatedly shown that the overnight dynamics
is qualitatively different from that during the day \cite{Chan,Edmonds,Zhong,Tsiakas,Chen,Lee,Gallo}.
Various reasons have been proposed for this: 
\begin{itemize}
\item A foreign equity which is mainly traded on some foreign market (that
is open during the night hours of the market studied) reflects mostly
its activity in their overnight volatility, and this activity might
be very different \cite{Chan} from the market under consideration. 
\item The majority of news relevant for fundamental stock price evaluation
(company profits, employment rates, general econometric forecasts,
wars and natural disasters, ...) are released overnight \cite{Corral},
and there exists a correlation between frequency of news releases
and volatilities \cite{Edmonds,Gallo}. 
\item While the market can react during the day to any outside perturbation,
it cannot do so during the night, which might also explain the higher
volatility immediately after the market opening \cite{Gallo}. 
\end{itemize}
Since research in economy is mostly driven by the hope for practical
applications, it is not surprising that the majority of the above
references were concerned with prediction. The general consensus seems
that overnight volatility is useful for predicting subsequent intra-day
volatility \cite{Blanc,Taylor,Chicheportiche}. This is an important
result. But prediction involves a model (GARCH \cite{Gallo,Chen},
SEMIFAR \cite{Chen}, or different versions of the stochastic volatility
model (SVM) \cite{Tsiakas,Zhong,Lee}), and none of the papers cited
above report \textit{model independent} analyses of the raw data themselves.
This is so in spite of the fact that data analyses not involving any
model and using only elementary methods and minimal assumptions would
be most useful for understanding the \textit{basic mechanism(s)} underlying
the phenomena.

It is the purpose of the present short note to provide just such an
elementary analysis. The methods used will be completely elementary,
and involve nothing more than (Spearman- \cite{Press}) cross correlations.
Yet the result is striking and completely unexpected, as far as significance
and robustness are concerned. We should point out that an extensive
statistical study of intra-day and overnight returns and volatilities
was recently made in \cite{Wang}, but since that analysis was not
guided by any theoretical considerations, the effect described below
was missed.

Let us use the index $k$ to count trading days (i.e., skipping weekend
and other non-trading days), and denote by $o_{k}$ and $c_{k}$ the
opening and closing prices of one particular equity. Intra-day log-returns
of this equity are defined as

\begin{equation}
d_{k}=\ln\frac{c_{k}}{o_{k}}\;,\label{eq:1}
\end{equation}
while overnight log-returns are

\begin{equation}
n_{k}=\ln\frac{o_{k}}{c_{k-1}}\;.\label{eq:2}
\end{equation}
Thus overnight returns are indexed by the index of the following day.
In case of weekends and holidays the over-``night'' returns include
all changes during the entire non-trading period. Volatilities are
in principle defined through the variances of log-returns as observed
over an extended time span. But when discussing them on a fine grained
temporal scale, they are usually replaced by the absolute values of
the log-returns (see e.g. footnote 11 in \cite{Tsiakas}). We will
follow this usage.

\begin{figure}
\begin{centering}
\includegraphics[scale=0.35]{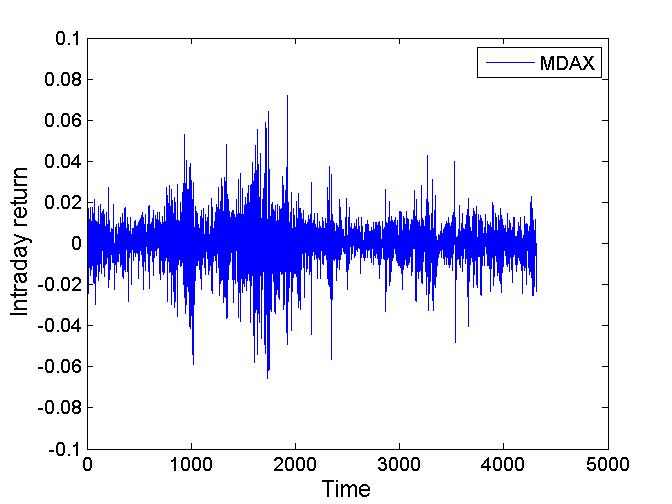} 
\includegraphics[scale=0.35]{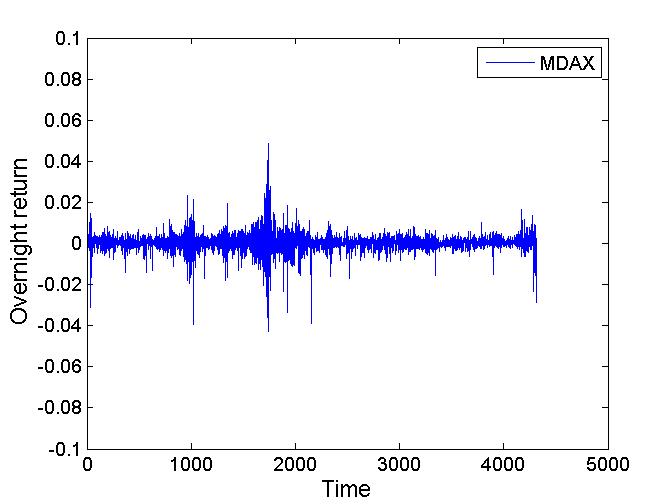}
\par\end{centering}

\protect\protect\caption{\label{fig:1}Historical time series of stock intra-day (panel a) and
overnight (panel b) returns for the MDAX index.}
\end{figure}

The data we studied consists of 21 individual stocks traded at various
stock exchanges (Exxon, Shell, General Electric, Ford, Goldmann-Sachs, Bank of America,
Citigroup, IBM, Microsoft, Cisco, AIG, BP, Caterpillar and Ford all traded at NYSE;
Siemens, Deutsche Bank, Lufthansa, VW and Bayer traded in Frankfurt; and Sony \& Mitsubishi 
traded in Tokyo) and 10 market indices and exchange-traded
funds (TecDax, MDax, DAX, Dow Jones, S\&P 100, Nasdaq, EuroSTOXX 50, SIM, S\&P/ASX and PowerShares QQQ). They were
mostly downloaded from Yahoo (https://finance.yahoo.com/), the rest
from finanzen.net (http://www.finanzen.net). The time sequences cover
between 10.4 and 45 years, with between 2612 and 13478 data points.
Before using them, we cleaned them from some of their artifacts (missing
data, wrong data, ...), but not of all. For instance, we did \textit{not}
remove jumps due to stock splitting. After cleaning, they show the
typical features well known from previous analyses, such as fat tails,
short-time correlations in the returns, and long-time correlations
in the volatilities. For typical examples, see Fig. 1. Notice that
these data still have outliers (mostly negative, due to crashes, bad
annual reports,...). The negative outliers occur mostly for the overnight
returns, consistent with the previous observation that negative news
are disseminated mostly when the markets are closed. The long autocorrelations
of the volatilities are seen both for daytime and overnight.

\begin{figure}
\begin{centering}
\includegraphics[scale=0.3]{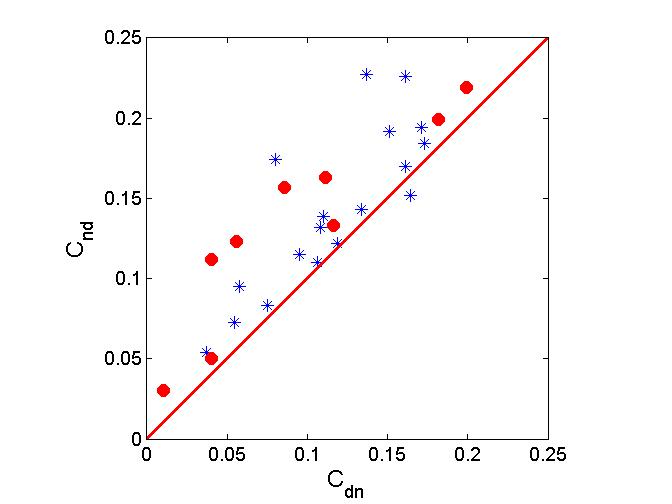} 
\par\end{centering}

\protect\protect\caption{Data for 31 equities. Each dot corresponds to one equity. 
The Spearman correlation between
intra-day volatilities and overnight volatilities during the \textit{subsequent}
night are plotted on the x-axis, while the correlations with the \textit{preceding}
night are on the y-axis.}
\label{corr-1.fig} 
\end{figure}

Our main concern is with cross-correlations between intra-day and
overnight volatilities. Due to the artifacts, irregularities, and strong
non-stationarity in the data, we did not use simple Pearson coefficients.
Instead we used Spearman coefficients \cite{Press}. Being based on
rank statistics, these are known to be much more robust. Indeed, the
results shown below would have been much less clearly visible had
we used Pearson coefficients. Alternatively we could also have used
Kendall's $\tau$ \cite{Kendall} or mutual informations \cite{Cover},
both of which are known to be similarly robust.

Our main results are shown in Figs.~2 and 3, In Fig.~\ref{corr-1.fig}
we show for each equity two cross-correlations between the ranks $r_{d_k}$
and $r_{n_k}$ of the two volatilities $|d_k|$ and $|n_k|$:

\begin{equation}
C_{nd}=\frac{\langle r_{d_k}r_{n_k} \rangle-\langle r_{d_k} \rangle\cdot\langle r_{n_k} \rangle}{\sigma_d\;\sigma_n}\label{eq:3}
\end{equation}

\noindent is the rank correlation between the intra-day volatility and the volatility
during the \textit{preceding} night ($\sigma_{d}$ and $\sigma_{n}$
are the square roots of the rank variances), while

\begin{equation}
C_{dn}=\frac{\langle r_{d_k} r_{n_{k+1}|}\rangle-\langle r_{d_k} \rangle\cdot\langle r_{n_k} \rangle}{\sigma_{d}\;\sigma_{n}}\label{eq:4}
\end{equation}
gives the analogous correlation with the \textit{following} night. We see that
in all cases

\begin{equation}
C_{nd}>C_{dn}.
\end{equation}
For some equities the difference is small, but for others it can be
more than a factor of two. In only one case the inequality was violated.
Thus the overnight volatility is much stronger correlated with the
volatility during the following day than during the preceding day.
Otherwise said, overnight volatilities seem to influence strongly
what goes on during the following trading day, but do not seem to
be strongly influenced by what was going on during the day before
\cite{footnote}.

\begin{figure}
\begin{centering}
\includegraphics[scale=0.25]{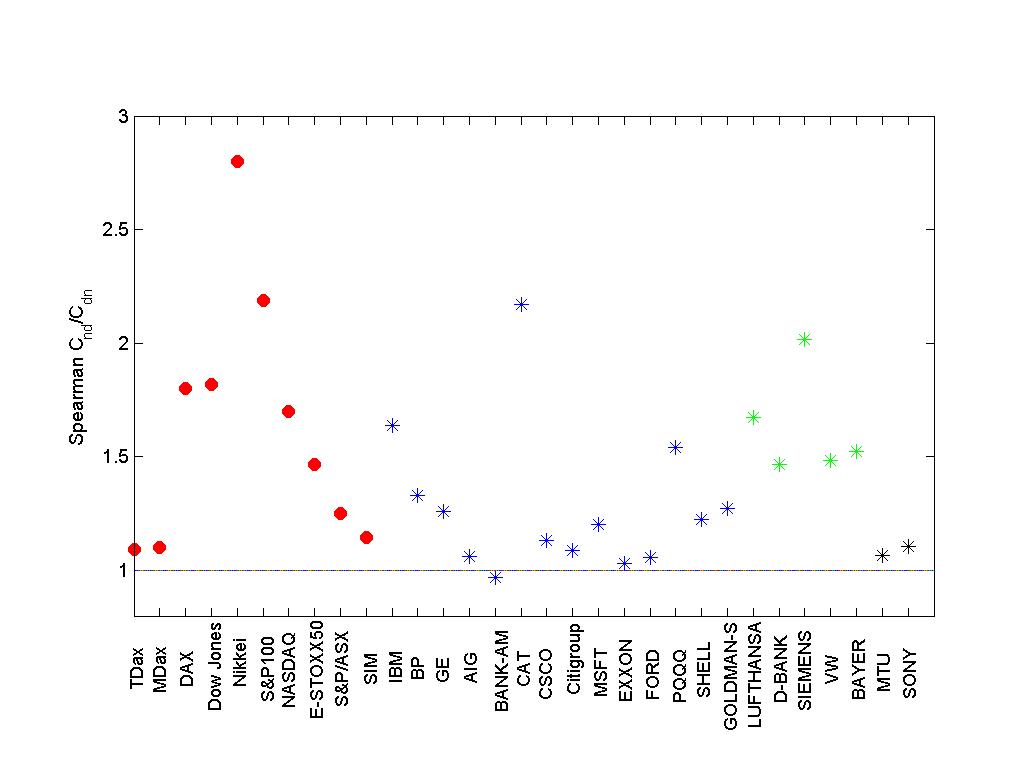} 
\par\end{centering}

\protect\caption{Ratios $C_{nd}$/ $C_{dn}$ for the 31 equities and
indices shown also in Fig.~\ref{corr-1.fig}. For stocks, the colors indicate the stock 
exchanges where they are traded.}
\label{corr-2.fig}
\end{figure}

The ratios $C_{nd}/C_{dn}$ for the equities used in Fig.~\ref{corr-1.fig}
are plotted also in Fig.~\ref{corr-2.fig}, where we have also specified
the equities. The first 10 entries in this figure are market indices,
while the others correspond to individual stocks. We see no big differences,
except that aggregated indices show a somewhat stronger effect. There
are also no noticeable differences related to the place where the equity
is traded, to the length of the time series, and -- in case of individual
stocks -- to the type of company.

At first sight this strong asymmetry looks very strange, in particular
since time asymmetry is usually considered to be very weak in financial
data. Many popular models (most noticeably all models of the ARCH family)
are time symmetric by construction, and where time asymmetry is seen
\cite{Zumbach,Blanc} it is only seen in very special observables.
But our findings are indeed compatible with previous analyses 
\cite{lockwood,Chan,Edmonds,Zhong,Tsiakas,Chen,Lee,Gallo}:
While the intra-day price dynamics is largely influenced by `chartist'
behavior, the overnight dynamics is mostly influenced by facts exogenous
to the stock market (or at least not directly related to the day-to-day
price evolution of the considered equity) and thus of `fundamentalist'
nature. What our results suggest is that `fundamentalist' information 
is more useful in prediction than `chartist' information.

The present analysis cannot of course specify which of the possible
external influences (foreign stock markets, company performance reports,
news about general economic indicators such as employment rates and
forecasted economic growth, wars, economic crises, natural disasters,
...) is of greatest importance for the overnight dynamics, but such
information could possibly be obtained by performing a larger study
similar to the present one in which equities are grouped according
to business sectors, stock exchanges, trading volume, bull \textit{versus}
bear markets, etc. Another improvement suggested by our analysis could
consist in replacing the simple cross correlations by partial correlations
or by transfer entropies \cite{Schreiber}, testing in this way for
linear or non-linear Granger causality \cite{Barnett}. It would be
of interest to see whether the asymmetry found in the present paper
is also present at larger time scales, by comparing day/night to night/day
results between more distant nights and days.

Finally, with the hindsight gained from this analysis, we might also
turn to signed returns (in contrast to volatilities) and test whether
some parts of a full 24 hours day have more influence on later periods
than others. The very fact that different regions in the phase space
of a recurrent system can have different powers of predictability
has been known for long time \cite{Farmer}.

We thank Andreas Kl\"umper for discussions and for carefully reading the 
manuscript, and Jean-Philippe Bouchaud and Spyros Skouras for comments. 
R.Z. thanks Holger Kantz and Peter Fulde for support and 
discussions. P.G. thanks Derek Belle for collaboration during an early 
stage of the work and Amer Shreim for discussions.


\begin{thebibliography}{10}
\bibitem{footnote} Notice in this context that the title of \cite{Zhong}
is wrong. That paper is concerned with the \textit{usefulness} of
overnight information for prediction, not with its \textit{predictability}.
\end{thebibliography}
\end{document}